# Accumulate-Repeat-Accumulate Codes: Systematic Codes Achieving the Binary Erasure Channel Capacity with Bounded Complexity

Henry D. Pfister
EPFL
School of Comp. and Comm. Science
Lausanne 1015, Switzerland
`henry.pfister@epfl.ch`

Igal Sason
Technion
Dept. of Electrical Engineering
Haifa 32000, Israel
`sason@ee.technion.ac.il`

**Abstract**

The paper introduces ensembles of accumulate-repeat-accumulate (ARA) codes which asymptotically achieve capacity on the binary erasure channel (BEC) with *bounded complexity* per information bit. It also introduces symmetry properties which play a central role in the construction of capacity-achieving ensembles for the BEC. The results here improve on the tradeoff between performance and complexity provided by the first capacity-achieving ensembles of irregular repeat-accumulate (IRA) codes with bounded complexity per information bit; these IRA ensembles were previously constructed by Pfister, Sason and Urbanke. The superiority of ARA codes with moderate to large block length is exemplified by computer simulations which compare their performance with those of previously reported capacity-achieving ensembles of LDPC and IRA codes. The ARA codes also have the advantage of being systematic.

*Index terms* – binary erasure channel (BEC), capacity, complexity, degree distribution (d.d.), density evolution (DE), iterative decoding, irregular repeat-accumulate (IRA) codes, systematic codes.

## 1 Introduction

Error correcting codes which employ iterative decoding algorithms are now considered state of the art in the field of low-complexity coding techniques. By now, there is a large collection of families of iteratively decoded codes including low-density parity-check (LDPC), turbo, repeat-accumulate and product codes; all of them demonstrate a rather small gap (in rate) to capacity with feasible complexity.

The study of capacity-achieving (c.a.) sequences of LDPC ensembles for the binary erasure channel (BEC) was initiated by Luby et al. [1] and Shokrollahi [2]. They show that it is possible to closely approach the capacity of an erasure channel with a simple iterative procedure whose complexity is linear in the block length of the code [1, 2]. Following these works, Oswald and Shokrollahi presented in [3] a systematic study of c.a. degree distributions for sequences of ensembles of LDPC codes whose transmission takes place over a BEC. Jin et al. introduced irregular repeat-accumulate (IRA) codes and presented a c.a. sequence of systematic IRA (SIRA) ensembles [4]. A new sequence of c.a.

SIRA codes with lower complexity was also introduced in [5]. All of the aforementioned codes have one major drawback; their decoding complexity scales like the log of the inverse of the gap (in rate) to capacity, which becomes unbounded as the gap to capacity vanishes [5, 6, 7].

In a previous paper [8], Pfister, Sason and Urbanke presented for the first time two sequences of ensembles of non-systematic IRA (NSIRA) codes which asymptotically (as their block length goes to infinity) achieve capacity on the BEC with bounded complexity per information bit. The new bounded complexity result in [8] is achieved by puncturing bits and allowing in this way a sufficient number of state nodes in the Tanner graph representing the codes. We note that for fixed complexity, the new codes in [8] eventually (for large enough block length) outperform any code proposed so far. However, the convergence speed to the ultimate performance limit happens to be quite slow, so for small to moderate block lengths, the new codes are not record breaking.

In this paper, we are interested in constructing c.a. codes on the BEC with bounded complexity per information bit which are also systematic codes. We would also like these codes to perform well at moderate block lengths and have low error floors. To this end, we make use of a new channel coding scheme, called "Accumulate-Repeat-Accumulate" (ARA) codes, which was recently introduced by Abbasfar, Divsalar, and Yao [9]. These codes are systematic and have both outstanding performance, as exemplified in [9, 10, 11], and a simple linear-time encoding. After defining an appropriate ensemble of irregular ARA codes, we construct a number of c.a. degree distributions. Simulations show that some of these ensembles perform quite well on the BEC at moderate block lengths. Therefore, we expect that irregular ARA codes optimized for general channels might also perform well at moderate block lengths.

Along the way, we study the symmetry of c.a. degree distributions and discover a new code structure which we call "Accumulate-LDPC" (ALDPC) codes. We show that c.a. degree distributions for this structure can be constructed easily based on the results of [8, Theorems 1 and 2]. This fact and structure was also proposed independently by Hsu and Anastasopoulos in [12].

## 2 Accumulate-Repeat-Accumulate Codes

In this section, we present our ensemble of ARA codes. Density evolution (DE) analysis of this ensemble is presented in the second part of this section using two different approaches which lead to the same equation for the fixed points of the iterative message-passing decoder (this equation will be called the "DE fixed point equation"). The connection between these two approaches is used later in this paper to state some symmetry properties which serve as an analytical tool for designing various c.a. ensembles for the BEC (e.g., ARA, IRA and ALDPC codes).

### 2.1 Description of ARA Codes

ARA codes can be viewed either as interleaved serially concatenated codes (i.e., turbo-like codes) or as sparse-graph codes (i.e., LDPC-like codes). From an encoding point of view, it is more natural to treat them as interleaved serially concatenated codes (see Fig. 1). Since their decoding algorithm is simply belief propagation on the appropriate Tanner graph (see Fig. 2), this leads one to view them also as sparse-graph codes.

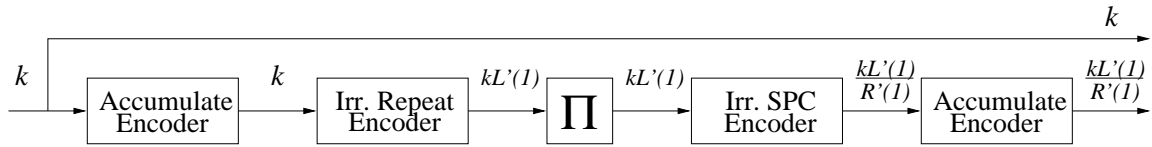

Figure 1: Block diagram for the systematic ARA ensemble ("Irr." and "SPC" stand for "irregular" and "single-parity check", respectively, and $\Pi$ stands for a bit interleaver.)

Treating these codes as sparse-graph codes also allows one to build large codes by "twisting" together many copies of a single small *protograph* [13, 14]. In general, this approach leads to very good codes and computationally efficient decoders.

In this work, we consider the ensemble of irregular ARA codes which is the natural generalization of irregular IRA codes [4, 8]. This ensemble differs slightly from those proposed in [9, 10, 11]. For this ensemble, we find that DE for the BEC can be computed in closed form and that algebraic methods can be used to construct c.a. sequences.

An irregular ensemble of ARA codes is defined by its degree distribution (d.d.). Nodes in the decoding graph will be referred to by the names given in Fig. 2. Let $L(x) = \sum_{i=1}^{\infty} L_i x^i$ be a power series where $L_i$ denotes the fraction of "punctured bit" nodes with degree $i$. Similarly, let $R(x) = \sum_{i=1}^{\infty} R_i x^i$ be a power series where $R_i$ denotes the fraction of "parity-check 2" nodes with degree $i$. In both cases, the degree refers only to the edges connecting the "punctured bit" nodes to the "parity-check 2" nodes. Similarly, let $\lambda(x) = \sum_{i=1}^{\infty} \lambda_i x^{i-1}$ and $\rho(x) = \sum_{i=1}^{\infty} \rho_i x^{i-1}$ form the d.d. pair from the edge perspective where $\lambda_i$ and $\rho_i$ designate the fraction of the edges which are connected to "punctured bit" nodes and "parity-check 2" nodes with degree $i$, respectively. We also assume that the permutation in Fig. 1 is chosen uniformly at random from the set of all permutations. The pair of degree distributions of an ARA ensemble is given by $(\lambda, \rho)$.

It is easy to show the following connections between the d.d. pairs w.r.t. the nodes and the edges in the graph:

$$\lambda(x) = \frac{L'(x)}{L'(1)}, \qquad \rho(x) = \frac{R'(x)}{R'(1)} \qquad (1)$$

or equivalently, since $L(0) = R(0) = 0$, then

$$L(x) = \frac{\int_0^x \lambda(t)\, dt}{\int_0^1 \lambda(t)\, dt}, \qquad R(x) = \frac{\int_0^x \rho(t)\, dt}{\int_0^1 \rho(t)\, dt}. \qquad (2)$$

## 2.2 Density Evolution of ARA Ensembles

We consider here the asymptotic analysis of ensembles of ARA codes. We assume that the codes are transmitted over a BEC with erasure probability $p$ and decoded with an iterative message-passing decoder.

A single decoding iteration consists of six smaller steps which are performed on the Tanner graph of Fig. 2. Messages are first passed downward from the "systematic bit" nodes through each layer to the "code bit" nodes. Then, messages are passed back upwards from the "code bit" nodes though each layer to the "systematic bit" nodes. Let $l$ designate the iteration number. Referring to Fig. 2, let $x_0^{(l)}$ and $x_5^{(l)}$ designate the probabilities of an erasure message from the "parity-check 1" nodes to the "punctured bit" nodes and vice-versa, let $x_1^{(l)}$ and $x_4^{(l)}$ be the probabilities of an erasure message from the "punctured bit" nodes to the "parity-check 2" nodes and vice versa, and finally, let $x_2^{(l)}$ and $x_3^{(l)}$ be the probabilities of an erasure message from the "parity-check 2" nodes to "code bit" nodes and vice versa. As the block length goes to infinity, the cycle-free condition holds with probability 1 and this implies that the messages become statistically independent with probability 1. Under this assumption, we obtain the following DE equations (from Fig. 2) for the message-passing iterative decoder:

$$
\begin{aligned}
x_0^{(l+1)} &= 1 - (1 - x_5^{(l)})(1-p) \\
x_1^{(l)} &= (x_0^{(l)})^2 \, \lambda(x_4^{(l)}) \\
x_2^{(l)} &= 1 - R(1 - x_1^{(l)}) \, (1 - x_3^{(l)}) \\
x_3^{(l)} &= p x_2^{(l)} \\
x_4^{(l)} &= 1 - (1 - x_3^{(l)})^2 \, \rho(1 - x_1^{(l)}) \\
x_5^{(l)} &= x_0^{(l)} \, L(x_4^{(l)})
\end{aligned}
$$

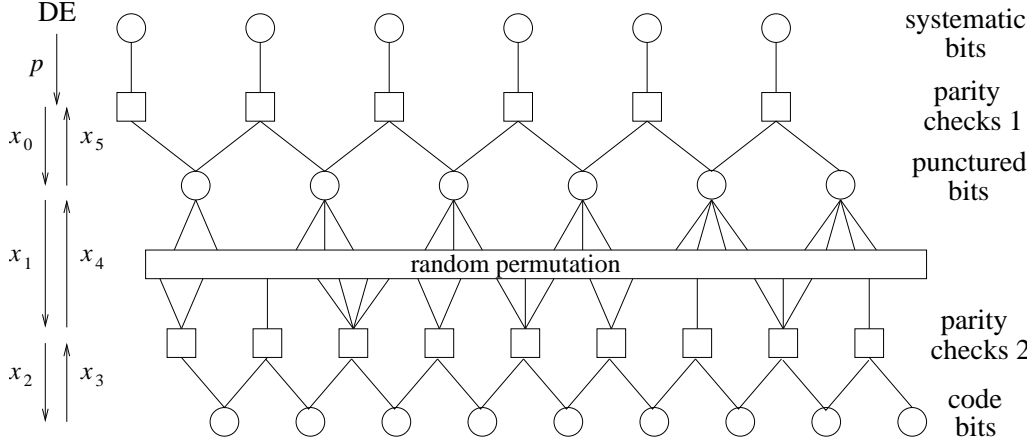

Figure 2: Tanner graph for the ARA ensemble.

A fixed point is implied by

$$\lim_{l \to \infty} x_i^{(l)} \triangleq x_i \qquad i = 1, \ldots, 5.$$

Algebra shows that if $x_1 \triangleq x$, then we obtain the following equation for the fixed points of the iterative decoder:

$$\frac{p^2 \, \lambda \left( 1 - \left( \frac{1-p}{1-pR(1-x)} \right)^2 \rho(1-x) \right)}{\left[ 1 - (1-p) \, L \left( 1 - \left( \frac{1-p}{1-pR(1-x)} \right)^2 \rho(1-x) \right) \right]^2} = x. \tag{3}$$

For ensembles of ARA codes whose transmission takes place over a BEC, the DE fixed point equation (3) can be also derived using a *graph reduction* approach.

We start by noting that any "code bit" node whose value is not erased by the BEC can be removed from the graph by merging its value into its two "parity-check 2" nodes. On the other hand, when the value of a "code bit" node is erased, one can merge the two

"parity-check 2" nodes which are connected to it (by summing the equations) and then remove the "code bit" node from the graph. This merging of the two "parity-check 2" nodes causes their degrees to be summed. Now, we consider the degree distribution (d.d.) of a single "parity-check 2" node in the reduced graph. This can be visualized as working from left to right in the graph, and assuming the value of the previous "code bit" node was known. The probability that there are $k$ erasures before the next observed "code bit" is given by $p^k(1-p)$. The graph reduction associated with this event causes the degrees of $k+1$ "parity-check 2" nodes (from the d.d. $R(x)$) to be summed. Therefore, the new d.d. of the "parity-check 2" nodes after the graph reduction is given by

$$\widetilde{R}(x) = \sum_{k=0}^{\infty} p^k(1-p)R(x)^{k+1} = \frac{(1-p)R(x)}{1-pR(x)}. \tag{4}$$

A similar graph reduction can be also performed on the "systematic bit" nodes in Fig. 2. Since degree 1 bit nodes (e.g., the "systematic bit" nodes in Fig. 2) only provide channel information, erasures make them worthless. So they can be removed along with their parity-checks (i.e., the "parity-check 1" nodes in Fig. 2) without affecting the decoder. On the other hand, whenever the value of a "systematic bit" node is observed (assume the value is zero w.o.l.o.g.), it can be removed leaving a degree 2 parity-check. Of course, degree 2 parity-checks imply equality and allow the connected "punctured bit" nodes to be merged (effectively summing their degrees). This gives a nice symmetry between the information bits and parity bits. Now, we consider the d.d. of a single "punctured bit" node in the reduced graph. This can be seen as working from left to right in the graph, and assuming the value of the previous "systematic bit" node was erased. The probability of the event where the values of $k$ "systematic bit" nodes are observed and the value of the next "systematic bit" node is erased by the channel is given by $(1-p)^k p$. The graph reduction associated with this event causes the degrees of $k+1$ "punctured bit" nodes (from the d.d. $L(x)$) to be summed. Therefore, the new d.d. of the "punctured bit" nodes after graph reduction is given by

$$\widetilde{L}(x) = \sum_{k=0}^{\infty}(1-p)^k pL(x)^{k+1} = \frac{pL(x)}{1-(1-p)L(x)}. \tag{5}$$

After the graph reduction, we are left with a standard LDPC code with new edge-perspective degree distributions given by

$$\widetilde{\lambda}(x) = \frac{\widetilde{L}'(x)}{\widetilde{L}'(1)} = \frac{p^2 \lambda(x)}{\big(1-(1-p)L(x)\big)^2} \tag{6}$$

$$\widetilde{\rho}(x) = \frac{\widetilde{R}'(x)}{\widetilde{R}'(1)} = \frac{(1-p)^2 \rho(x)}{\big(1-pR(x)\big)^2}. \tag{7}$$

After the aforementioned graph reduction, all the "systematic bit" nodes and "code bit" nodes are removed. Therefore the residual LDPC code effectively sees a BEC whose erasure probability is 1, and the DE fixed point equation is given by

$$\widetilde{\lambda}\big(1-\widetilde{\rho}(1-x)\big) = x. \tag{8}$$

Based on (6) and (7), the last equation is equivalent to (3). We note that although $\widetilde{\lambda}$ and $\widetilde{\rho}$ which are given in (6) and (7), depend on the erasure probability of the BEC ($p$), for simplicity of notation, we do not write this dependency explicitly in our notation. However, in Section 3, when discussing symmetry properties and replacing $p$ by $1-p$, the erasure probability is written explicitly in these tilted degree distributions.

# 3 Symmetry Properties of Capacity-Achieving Codes

In this section, we discuss the symmetry between the bit and check degree distributions of c.a. ensembles for the BEC. First, we describe this relationship for LDPC codes, and then we extend it to ARA codes. The extension is based on analyzing the decoding of ARA codes in terms of graph reduction and the DE analysis of LDPC codes.

## 3.1 Symmetry Properties of Capacity-Achieving LDPC Codes

The relationship between the bit d.d. and check d.d. of c.a. ensembles of LDPC codes can be expressed in a number of ways. Starting with the DE fixed point equation

$$p\lambda\bigl(1 - \rho(1 - x)\bigr) = x \tag{9}$$

where $p$ designates the erasure probability of the BEC, we see that picking either the d.d. $\lambda$ or $\rho$ determines the other d.d. exactly. In this section, we make this notion precise and use it to expose some of the symmetries of c.a. LDPC codes.

A few definitions are needed to discuss things properly. Following the notation in [3], let $\mathcal{P}$ be the set of d.d. functions (i.e., functions $f$ with non-negative power series expansions around zero which satisfy $f(0) = 0$ and $f(1) = 1$); this set is defined by

$$\mathcal{P} \triangleq \left\{ f : f(x) = \sum_{k=1}^{\infty} f_k x^k,\ x \in [0,1],\quad f_k \geq 0,\ f(0) = 0,\ f(1) = 1 \right\}.$$

Let $\mathcal{T}$ be an operator which transforms invertible functions $f : [0,1] \to [0,1]$ according to the rule

$$\mathcal{T}f(x) \triangleq 1 - f^{-1}(1 - x)$$

where $f^{-1}$ is the inverse function of $f$. The function $\mathcal{T}f$ is well-defined on $[0,1]$ for any function $f$ which is strictly monotonic on this interval, and therefore for any function in $\mathcal{P}$. We will say that two d.d. functions $f$ and $g$ are *matched* if $\mathcal{T}f = g$ (since $\mathcal{T}^2 f \equiv f$, the equality $\mathcal{T}f = g$ implies that $\mathcal{T}g = f$). Finally, let $\mathcal{A}$ be the set of all functions $f \in \mathcal{P}$ such that $\mathcal{T}f \in \mathcal{P}$, i.e.,

$$\mathcal{A} \triangleq \bigl\{ f : f \in \mathcal{P},\ \mathcal{T}f \in \mathcal{P} \bigr\}.$$

The connection with LDPC codes is that finding some $f \in \mathcal{A}$ is typically the first step towards proving that $(f, \mathcal{T}f)$ is a c.a. d.d. pair. Truncation and normalization issues which depend on the erasure probability of the BEC must also be considered. When $p = 1$, many of these issues disappear, so we denote the set of d.d. pairs which satisfy (9) by

$$\begin{aligned}
\mathcal{C}_{\text{LDPC}} &\triangleq \bigl\{ (\lambda, \rho) \in \mathcal{P} \times \mathcal{P} \mid \lambda\bigl(1 - \rho(1-x)\bigr) = x \bigr\} \\
&= \bigl\{ (\lambda, \rho) \mid \lambda \in \mathcal{A},\ \rho = \mathcal{T}\lambda \bigr\}.
\end{aligned}$$

The *symmetry property* of c.a. LDPC codes (with rate 0) asserts that

$$(\lambda, \rho) \in \mathcal{C}_{\text{LDPC}} \xleftrightarrow{\text{symmetry}} (\rho, \lambda) \in \mathcal{C}_{\text{LDPC}}. \tag{10}$$

One can prove this result by transforming (9) when $p = 1$. First, we let $x = 1 - \rho^{-1}(1 - y)$, which gives $\lambda(y) = 1 - \rho^{-1}(1 - y)$, then we rewrite this expression as $\rho\bigl(1 - \lambda(y)\bigr) = 1 - y$ and finally, let $y = 1 - z$ to get $\rho\bigl(1 - \lambda(1 - z)\bigr) = z$. Comparing this with the DE fixed point equation (9) when $p = 1$ shows the symmetry between $\lambda$ and $\rho$.

## 3.2 Symmetry Properties of ARA Codes

The decoding of an ARA code can be broken into two stages. The first stage transforms the ARA code into an equivalent LDPC code via graph reduction, and the second stage decodes the LDPC code. This allows us to describe the symmetry property of c.a. ARA codes in terms of the symmetry property of c.a. LDPC codes. For $f \in \mathcal{P}$, let us define

$$\widetilde{f}_p(x) \triangleq \frac{(1-p)^2 f(x)}{\left(1 - \frac{p \int_0^x f(t)dt}{\int_0^1 f(t)dt}\right)^2}. \tag{11}$$

One can write the d.d. pair $(\widetilde{\lambda}, \widetilde{\rho})$ after graph reduction by combining (2), (6) and (7) which gives

$$\widetilde{\lambda} = \widetilde{\lambda}_{1-p}, \quad \widetilde{\rho} = \widetilde{\rho}_p.$$

This allows graph reduction to be interpreted as a mapping $\mathcal{G}_{\mathrm{ARA}}$ from an ARA d.d. pair to an LDPC d.d. pair which can be expressed as

$$(\lambda, \rho) \xleftarrow{\mathcal{G}_{\mathrm{ARA}}} (\widetilde{\lambda}_{1-p}, \widetilde{\rho}_p).$$

The inverse of the graph reduction mapping is represented by a dashed arrow because this inverse mapping, while always well-defined, does not necessarily preserve the non-negativity of d.d. functions.

Referring to ensembles of ARA codes, the set of d.d. pairs which satisfy the DE fixed point equation (3) is given by

$$\mathcal{C}_{\mathrm{ARA}}(p) \triangleq \left\{ (\lambda, \rho) \in \mathcal{P} \times \mathcal{P} \mid \widetilde{\lambda}_{1-p}\big(1 - \widetilde{\rho}_p(1-x)\big) = x \right\}$$

where the equivalence to (3) follows from (6), (7) and (11).

The symmetry between the bit and check degree distributions of a c.a. ARA ensemble follows from the symmetry relationship in (10), and the equivalence between a d.d. pair $(\lambda, \rho)$ for ARA codes and the d.d. pair $(\widetilde{\lambda}_{1-p}, \widetilde{\rho}_p)$ for LDPC codes.

The complete symmetry relationship is therefore given in the following diagram:

$$
\begin{array}{ccc}
(\lambda, \rho) \in \mathcal{C}_{\mathrm{ARA}}(p) & \xleftrightarrow{\text{ARA symmetry}} & (\rho, \lambda) \in \mathcal{C}_{\mathrm{ARA}}(1-p) \\
\mathcal{G}_{\mathrm{ARA}} \updownarrow & & \updownarrow \mathcal{G}_{\mathrm{ARA}} \\
(\widetilde{\lambda}_{1-p}, \widetilde{\rho}_p) \in \mathcal{C}_{\mathrm{LDPC}} & \xleftrightarrow{\text{LDPC symmetry}} & (\widetilde{\rho}_p, \widetilde{\lambda}_{1-p}) \in \mathcal{C}_{\mathrm{LDPC}}
\end{array}
$$

The inverse of the graph reduction mapping is represented by the dashed arrow because this inverse transformation is only valid if it is known ahead of time that the power series expansions of $\lambda$ and $\rho$ are non-negative. It turns out that this symmetry is very useful in order to generate new d.d. pairs which satisfy the DE equality in (8). An alternative way to show this symmetry explicitly is rewriting (8)

$$\widetilde{\lambda}_{1-p}\big(1 - \widetilde{\rho}_p(x)\big) = x$$

and using the symmetry property (10) for LDPC codes to rewrite it as

$$\widetilde{\rho}_p\big(1 - \widetilde{\lambda}_{1-p}(x)\big) = 1 - x.$$

From (6) and (7), the expansion of the last equation gives

$$\frac{(1-p)^2 \, \rho\left(1 - \frac{p^2 \lambda(x)}{\left(1-(1-p)L(x)\right)^2}\right)}{\left(1 - p\, R\left(1 - \frac{p^2 \lambda(x)}{\left(1-(1-p)L(x)\right)^2}\right)\right)^2} = 1 - x. \tag{12}$$

Since the swapping $L(x) \leftrightarrow R(x)$, $\lambda(x) \leftrightarrow \rho(x)$, $p \leftrightarrow 1-p$, and $x \leftrightarrow 1-x$ maps this equation back to (3), then we can take any d.d. pair $(\lambda, \rho)$ which satisfies (3) for $p = p^*$ and swap $\lambda$ with $\rho$ (and hence, $L$ and $R$ are also swapped) to get a new d.d. pair which satisfies (12) for $p = 1 - p^*$ (equations (3) and (12) should be satisfied for all $x \in [0,1]$, so switching between $x$ and $1-x$ has no relevance).

## 3.3 Symmetry Properties of NSIRA Codes

Now, we consider the graph reduction process and symmetry properties of non-systematic irregular repeat-accumulate (NSIRA) codes (for preliminary material on NSIRA codes, the reader is referred to [8, Section 2]). In this respect, we introduce a new ensemble of codes which we call "Accumulate-LDPC" (ALDPC) codes. These codes are the natural image of NSIRA codes under the symmetry transformation. In fact, this ensemble was discovered by applying the symmetry transformation to previously known c.a. code ensembles. Their decoding graph can be constructed from the ARA decoding graph (see Fig. 2) by removing bottom accumulate structure.

Since an NSIRA code has no accumulate structure attached to the "punctured bit" nodes, the graph reduction process affects only the d.d. of the "parity-check 2" nodes. Therefore, graph reduction acts as a mapping $\mathcal{G}_{\text{NSIRA}}$ from the NSIRA d.d. pair $(\lambda, \rho)$ to the LDPC d.d. pair $(\lambda, \widetilde{\rho}_p)$. This yields that for ensembles of NSIRA codes, the set of d.d. pairs which satisfy the DE fixed point equation is given by

$$\mathcal{C}_{\text{NSIRA}}(p) \triangleq \left\{ (\lambda, \rho) \in \mathcal{P} \times \mathcal{P} \mid \lambda\bigl(1 - \widetilde{\rho}_p(1-x)\bigr) = x \right\}.$$

An ALDPC code has no accumulate structure attached to the "parity-check 2" nodes, and therefore the graph reduction process only affects the d.d. of the "punctured bit" nodes. Hence, graph reduction acts as a mapping $\mathcal{G}_{\text{ALDPC}}$ from the ALDPC d.d. pair $(\lambda, \rho)$ to the LDPC d.d. pair $(\widetilde{\lambda}_{1-p}, \rho)$. For ALDPC ensembles, the set of d.d. pairs which satisfy the DE fixed point equation is therefore given by

$$\mathcal{C}_{\text{ALDPC}}(p) \triangleq \left\{ (\lambda, \rho) \in \mathcal{P} \times \mathcal{P} \mid \widetilde{\lambda}_{1-p}\bigl(1 - \rho(1-x)\bigr) = x \right\}.$$

The symmetry between NSIRA and ALDPC ensembles follows from the symmetry relationship in (10), the equivalence between a d.d. pair $(\lambda, \rho)$ for NSIRA codes and the d.d. pair $(\lambda, \widetilde{\rho}_p)$ for LDPC codes, and the relationship between a d.d. pair $(\lambda, \rho)$ for ALDPC codes and the d.d. pair $(\widetilde{\lambda}_{1-p}, \rho)$.

The symmetry relationship is therefore given in the following diagram.

$$\begin{array}{ccc} (\lambda, \rho) \in \mathcal{C}_{\text{NSIRA}}(p) & \xleftrightarrow{\text{symmetry}} & (\rho, \lambda) \in \mathcal{C}_{\text{ALDPC}}(1-p) \\ \mathcal{G}_{\text{NSIRA}} \Big\updownarrow & & \Big\updownarrow \mathcal{G}_{\text{ALDPC}} \\ (\lambda, \widetilde{\rho}_p) \in \mathcal{C}_{\text{LDPC}} & \xleftrightarrow{\text{LDPC symmetry}} & (\widetilde{\rho}_p, \lambda) \in \mathcal{C}_{\text{LDPC}} \end{array}$$

As before, the inverse of each graph reduction mapping is represented by a dashed arrow because this inverse transformation is only valid if it is known ahead of time that the power series expansions of $\lambda$ and $\rho$ are non-negative.

## 3.4 Connections with Forney's Transform

In [15], Forney introduces a graph transformation which maps the factor graph of any group code to the factor graph of the dual group code. For factor graphs of binary linear codes which only have equality and parity constraints (i.e., no trellis constraints), this operation is equivalent to swapping equality and parity constraints (e.g., bit nodes and check nodes). Forney's approach represents observations by half-edges, and these remain attached to the original node even though the nature of that node has changed. For example, Forney's transform maps an LDPC code with parity-check matrix $H$ to a low-density generator-matrix (LDGM) code with generator matrix $H$ and the half-edges attached to the bit nodes of the LDPC code are attached to the parity-check nodes of the LDGM code.

Using Forney's transform, we see that the swapping of $\lambda$ and $\rho$ described by our symmetry mappings actually transforms the original code ensemble into the dual code ensemble. Let the design rate of the original ensemble be $R$, then the design rate of the dual ensemble is $1 - R$. This means that if we want to have any chance of achieving capacity, we must also map the channel erasure probability $p$ to $1 - p$. Therefore, our symmetry relationships show that ARA, NSIRA, and ALDPC ensembles which are c.a. on BEC under iterative decoding also have dual ensembles which are c.a. on the BEC under iterative decoding.[1]

Finally, we note that the basic structure of ARA codes is preserved under Forney's transform. In particular, this means that we can construct self-dual ARA codes (i.e., rate $\frac{1}{2}$) by choosing the square connection matrix between the "punctured bit" nodes and the "parity-check 2" nodes to be symmetric.

## 4 Bit-Regular and Check-Regular Capacity-Achieving Ensembles with Bounded Complexity for the BEC

This section gives explicit constructions of c.a. ARA ensembles for the BEC, where these ensembles are either bit-regular or check-regular. As will be observed, these ensembles possess bounded complexity (per information bit) as the gap to capacity vanishes.

The symmetry property in Section 3.2 allows one for example to design an ensemble of high rate ARA codes, and get automatically (by switching between the pair of degree distributions) a new ensemble of ARA codes which is suited for low rate applications. We will rely on this symmetry property in Section 4.2 when we transform a bit-regular ARA ensemble designed for a BEC with erasure probability $p \in (0, p^*]$ into a check-regular ensemble designed for $p \in [1 - p^*, 1)$. We also rely on the fact that the method in Section 4.1 for computing the function $R$ given the function $L$ can be easily inverted using the symmetry property. This means that given an algorithm to solve for $R(x)$ in terms of $L(x)$ for a certain $p_0$, the inverse algorithm which solves $L(x)$ in terms of $R(x)$ is exactly the same, except that $p_0$ is replaced by $1 - p_0$.

---

[1]To be precise, we actually need to consider sequences on ensembles which are c.a. and relate them to sequences of dual ensembles. This distinction is rather cumbersome and does not cause problems in this case.

## 4.1 Solving for $R(x)$ in terms of $L(x)$

Given $L(x)$, we start with the calculation of $\lambda(x) = \frac{L'(x)}{L'(1)}$. Then $\widetilde{\lambda}(x)$ is calculated from (6), and $\widetilde{\rho}(x) = 1 - \widetilde{\lambda}^{-1}(1 - x)$ is calculated from (8). Further algebra gives

$$\rho(x) = \frac{\widetilde{\rho}(x)}{\big(1 - p + pQ(x)\big)^2}, \qquad Q(x) \triangleq \frac{\int_0^x \widetilde{\rho}(t)\,dt}{\int_0^1 \widetilde{\rho}(t)\,dt}. \tag{13}$$

As long as we have $\widetilde{\rho}(1) = 1$, then evaluating (13) at $x = 1$ gives $\rho(1) = 1$. Therefore, there is no need to truncate the power series of $\rho$. As we noted above, a very similar approach can be applied to solve for $L(x)$ in terms of $R(x)$; due to the symmetry property, one can simply apply the above procedure to a parity-check d.d. $R(x)$ with an erasure probability of $1 - p$.

## 4.2 Capacity-Achieving ARA Ensembles

The relationship between bit-regular and check-regular c.a. ensembles of ARA codes follows from the symmetry properties presented in Section 3.2, so we choose to focus on a bit-regular ARA ensemble. Let $\lambda(x) = x^2$, so $L(x) = x^3$, and from (6)

$$\widetilde{\lambda}(x) = \frac{p^2 x^2}{\big(1 - (1 - p)x^3\big)^2}.$$

Based on (8), we get

$$\widetilde{\rho}^{-1}(x) = 1 - \widetilde{\lambda}(1 - x)$$
$$= 1 - \frac{p^2(1-x)^2}{\big(1 - (1-p)(1-x)^3\big)^2}.$$

This is exactly [8, Eq. (39)] with $p$ replaced by $1 - p$ and the d.d. $\rho$ switched with the d.d. $\lambda$. Therefore, we obtain from [8, Theorem 2] that the tilted d.d. $\widetilde{\rho}$ gets the form

$$\widetilde{\rho}(x) = 1 + \frac{2(1-p)(1-x)^2 \sin\left(\frac{1}{3} \arcsin\left(\sqrt{-\frac{27(1-p)(1-x)^{\frac{3}{2}}}{4p^3}}\right)\right)}{\sqrt{3}\, p^4 \left(-\frac{(1-p)(1-x)^{\frac{3}{2}}}{p^3}\right)^{\frac{3}{2}}}. \tag{14}$$

This allows one to write the d.d. $\rho$ compactly using (13). It was verified numerically that for $p \leq 0.384$, the first 300 coefficients in the power series expansion of $R(x)$ are positive. It also holds in general that any d.d. pair satisfying (3) has a design rate equal to the capacity of the BEC. Therefore, it appears that the d.d. pair above characterizes a c.a. ensemble of bit-regular ARA codes over the BEC; the capacity of the BEC is achieved with bounded complexity for rates greater than 0.616. Using the symmetry between $\widetilde{\lambda}$ and $\widetilde{\rho}$ (see Section 3), this also implies that for rates less than 0.384, the ensemble of check-regular ARA codes with $R(x) = x^3$ achieves capacity over the BEC with bounded complexity. We note that the convergence speed of the d.d. for the parity-check nodes is relatively fast. As an example, for $p = 0.3$, the fraction of check nodes with degree less than 32 is equal to 0.968.

## 4.3 Capacity-Achieving ALDPC Ensembles

Using the symmetry relationship between NSIRA and ALDPC ensembles from Section 3.3, we find that we already have from [8, Theorems 1 and 2] two c.a. ensembles of ALDPC codes. These ensembles are based on the bit-regular and check-regular NSIRA ensembles of [8]. This was also observed independently by Hsu and Anastasopoulos [12].

Using symmetry, the check-regular NSIRA ensemble gives a bit-regular ALDPC ensemble which provably achieves capacity with bounded complexity for $p \in [0.05, 1)$. Random puncturing can be viewed as increasing the effective erasure rate of the channel, and therefore puncturing extends this range to $(0, 1)$. Likewise, using symmetry, the bit-regular NSIRA ensemble gives a check-regular ALDPC ensemble which provably achieves capacity with bounded complexity for $p \in \left[\frac{12}{13}, 1\right)$. Again, random puncturing can be used to extend the valid range to $(0, 1)$.

# 5  Capacity-Achieving Ensembles with Bounded Complexity: Constructions Based on LDPC Codes

In this section, we introduce another way of constructing c.a. ensembles of ARA codes for the BEC. Rather then solving for the function $R$ in terms of the function $L$ (as in Section 4.1) or doing the inverse via the symmetry property, we consider here another natural way of searching for c.a. degree distributions. We start by choosing a candidate d.d. pair $(\widetilde{\lambda}, \widetilde{\rho})$ which satisfies equation (8) and test to see if it can be used to construct an ensemble of c.a. ARA codes. The testing process starts by mapping the tilted pair $(\widetilde{\lambda}, \widetilde{\rho})$ back to $(\lambda, \rho)$ via (6) and (7), and then testing the non-negativity of the resulting power series of $\lambda$ and $\rho$.

Following the notation in Section 3.1, it enables one to rewrite (8) as $\widetilde{\rho} = \mathcal{T}\widetilde{\lambda}$ (so the tilted degree distributions $\widetilde{\lambda}$ and $\widetilde{\rho}$ are matched), and gives a compact description of capacity-achieving d.d. pairs of LDPC codes. We note that since $\mathcal{T}^2 f = f$ for an arbitrary function $f$ which has an inverse, then $f \in \mathcal{A}$ if and only if $\mathcal{T}f \in \mathcal{A}$. Based on (8), we obtain that we need to choose the tilted d.d. so that $\widetilde{\lambda} \in \mathcal{P}$ and also $\mathcal{T}\widetilde{\lambda} \in \mathcal{P}$, i.e., we need that the d.d. $\widetilde{\lambda}$ (or $\widetilde{\rho}$) both belong to the set $\mathcal{A}$. The reader is referred to [3, Lemma 1] which considers basic properties of the set $\mathcal{A}$ and the transformation $\mathcal{T}$.

So far, by choosing $\widetilde{\lambda} \in \mathcal{A}$ (or $\widetilde{\rho} \in \mathcal{A}$), we only know that both tilted d.d. have non-negative power series expansions. This property does not ensure that both of the original (i.e., non-tilted) d.d. $\lambda$ and $\rho$ also have non-negative power series expansions. Calculation of $\lambda$ and $\rho$ from the tilted d.d. $\widetilde{\lambda}$ and $\widetilde{\rho}$ is not straightforward since both equations involve the d.d. $L$ and $R$ which are the normalized integrals of the unknown $\lambda$ and $\rho$. In order to overcome this difficulty in solving the two integral equations, we suggest calculating the tilted d.d. pair w.r.t. the nodes of the graph using

$$\widetilde{L}(x) = \frac{\int_0^x \widetilde{\lambda}(t)\, dt}{\int_0^1 \widetilde{\lambda}(t)\, dt}, \quad \widetilde{R}(x) = \frac{\int_0^x \widetilde{\rho}(t)\, dt}{\int_0^1 \widetilde{\rho}(t)\, dt}. \tag{15}$$

The original d.d. pair w.r.t. the nodes (i.e., the original d.d. pair before the graph reduction) can be calculated from Eqs. (4) and (5). We obtain that

$$L(x) = \frac{\widetilde{L}(x)}{p + (1-p)\widetilde{L}(x)}, \quad R(x) = \frac{\widetilde{R}(x)}{1 - p + p\widetilde{R}(x)} \tag{16}$$

and then use equation (1) to find $(\lambda, \rho)$. The critical issue here is to verify whether the functions $L$ and $R$ have non-negative power series expansions.

## 5.1 Capacity-Achieving ARA Ensembles

It is easy to verify that the function
$$f(x) = \frac{(1-b)x}{1-bx}, \quad 0 < b < 1 \tag{17}$$
belongs to the set $\mathcal{A}$ and also $\mathcal{T}f = f$; in the case where $\mathcal{T}f = f$, the function $f$ is said to be matched to itself. Therefore, based on (8), we examine here whether the choice $\widetilde{\lambda}(x) = \widetilde{\rho}(x) = \frac{(1-b)x}{1-bx}$ can be transformed into an ensemble of ARA codes whose degree distributions have non-negative power series expansions. From (15) and (16), we get
$$\widetilde{L}(x) = \widetilde{R}(x) = \frac{bx + \ln(1-bx)}{b + \ln(1-b)} \tag{18}$$
and
$$L(x) = \frac{bx + \ln(1-bx)}{p\left[b + \ln(1-b)\right] + (1-p)\left[bx + \ln(1-bx)\right]} \tag{19}$$
$$R(x) = \frac{bx + \ln(1-bx)}{(1-p)\left[b + \ln(1-b)\right] + p\left[bx + \ln(1-bx)\right]}. \tag{20}$$
The asymptotic behavior of the d.d. pairs w.r.t. the nodes and the edges is given by
$$L_k, R_k = O\left(\frac{b^k}{k \ln^2(k)}\right), \quad \lambda_k, \rho_k = O\left(\frac{b^k}{\ln^2(k)}\right).$$
We believe the performance advantage of this ensemble over other c.a. ensembles is mainly due to the *exponential* decay of the d.d. coefficients.

It has been observed empirically, that the power series expansions of both $R$ and $L$ are non-negative if and only if $p$ satisfies the inequality
$$\frac{1}{1 - \frac{13-\sqrt{61}}{9}\left(b + \ln(1-b)\right)} \leq p \leq 1 - \frac{1}{1 - \frac{13-\sqrt{61}}{9}\left(b + \ln(1-b)\right)} \tag{21}$$
and
$$b \in [b^*, 1), \quad b^* \triangleq W(-e^{-\frac{25+\sqrt{61}}{12}}) + 1 \approx 0.9304$$
where $W$ designates the Lambert W-function (i.e., $w = W(x)$ is the solution to the equation $we^w = x$, which is a real number for $x > -\frac{1}{e}$). These conditions follow since it was observed numerically that the strongest conditions for the non-negativity of the power series expansions of $L$ and $R$ are implied by the coefficients $L_6$ and $R_6$, respectively.

The encoding and decoding complexities per information bit of the considered c.a. ensembles of ARA codes for the BEC are bounded and given by
$$\chi_{\text{E}}, \chi_{\text{D}} = \frac{3-p}{1-p} - \frac{b^2 p}{(1-b)[b + \ln(1-b)]}.$$
For fixed $p$, the complexity is a monotonic increasing function of $b$ (which becomes unbounded as $b \to 1^-$). To minimize the encoding/decoding complexity, we wish to choose the smallest value of $b$ in the interval $(0, 1)$ so that all the coefficients in the power series expansions of the d.d. pair $L$ and $R$ are non-negative. We therefore need to choose the smallest value of $b$ ($0 < b < 1$) which satisfies the condition in (21). Since $p$ and $1-p$ imply the same value of $b$ in (21), the required value of $b$ is given by
$$b = W(-e^{-1-a}) + 1, \quad a \triangleq \frac{13 + \sqrt{61}}{12}\left(\frac{1}{\min(p, 1-p)} - 1\right).$$

## 5.2 Capacity-Achieving NSIRA Ensembles

In this section, we construct ensembles of NSIRA codes using LDPC codes whose degree distributions from the edge perspective are matched. We apply here the concept of DE via graph reduction to ensembles of uniformly interleaved NSIRA codes. In this case, the graph reduction only applies to the "parity-check 2" nodes (see Fig. 2). This is because the upper part of Fig. 2 does not exist in the Tanner graph of NSIRA codes (i.e., the "punctured bit" nodes in this figure are the "information bit" nodes in the graph of NSIRA codes). Based on graph reduction, we obtain that $L = \widetilde{L}$ for ensembles of NSIRA codes, while the functions $R$ and $\widetilde{R}$ satisfy the equality in (16). In a similar manner, the equality $\lambda = \widetilde{\lambda}$ holds for NSIRA ensembles while equality (7) is satisfied for the degree distributions of the parity-checks from the edge perspective. We note that from (7) and (8), the fixed point of the DE equations for NSIRA ensembles is given by

$$\lambda\left(1 - \frac{(1-p)^2 \rho(1-x)}{\left(1 - pR(1-x)\right)^2}\right) = x.$$

This equation coincides with [8, Eq. (6)] (where $x_0$ is replaced by $x$).

For the construction of ensembles of NSIRA codes using LDPC codes whose degree distributions from the edge perspective are both matched to themselves, we rely as a starting point on the function $f$ in (17) which forms a d.d. which is matched to itself, and set $\widetilde{\lambda}(x) = \widetilde{\rho}(x) = \frac{(1-b)x}{1-bx}$ for $0 < b < 1$, similarly to Section 5.1. For the considered ensemble of NSIRA codes, the d.d. $L(x)$ is then equal to $\widetilde{L}(x)$ in (18), i.e.,

$$L(x) = \frac{bx + \ln(1-bx)}{b + \ln(1-b)}.$$

From this, we see that there are no degree-1 "information bit" nodes, and that the fraction of "information bit" nodes with degree $i$ is given by

$$L_i = -\frac{b^i}{i} \frac{1}{b + \ln(1-b)}, \quad i = 2, 3, \ldots.$$

The non-negativity of the sequence $\{L_i\}$ holds when $0 < b < 1$ (so $b + \ln(1-b) < 0$). For the NSIRA ensemble considered in this example, there is no requirement on the erasure probability $p$ for keeping the power series expansion of the d.d. $L$ to be non-negative. It has been empirically observed that the following condition on $p$ needs to be satisfied so that the power series expansion of the d.d. $R$ will be non-negative:

$$p \leq 1 - \frac{1}{1 - \frac{13 - \sqrt{61}}{9} [b + \ln(1-b)]}. \tag{22}$$

By comparing it to the parallel requirement for the ARA ensemble, as given in (21), one observes that (22) requires a weaker condition on $p$ which is only the upper bound on $p$ in (21). As mentioned above, the d.d. $R$ is the same as for the ARA ensemble in Section 5.1. The encoding and decoding complexities of this ensemble are equal and have the form

$$\chi_{\mathrm{E}} = \chi_{\mathrm{D}} = \frac{2}{1-p} - \frac{b^2}{(1-b)[b + \ln(1-b)]}.$$

This gives an explicit construction of NSIRA ensembles from LDPC codes whose degree distributions from the edge perspective are matched to themselves. In general, we find by computer simulations for finite-length codes over the BEC that ARA codes have the best performance.

# Acknowledgment

The authors wish to acknowledge Rüdiger Urbanke for stimulating discussions during the preparation of the work in [8] which motivated the research in this paper. This work was supported by a research grant from Intel Israel. The work of I. Sason was also supported by the Taub and Shalom Foundations.

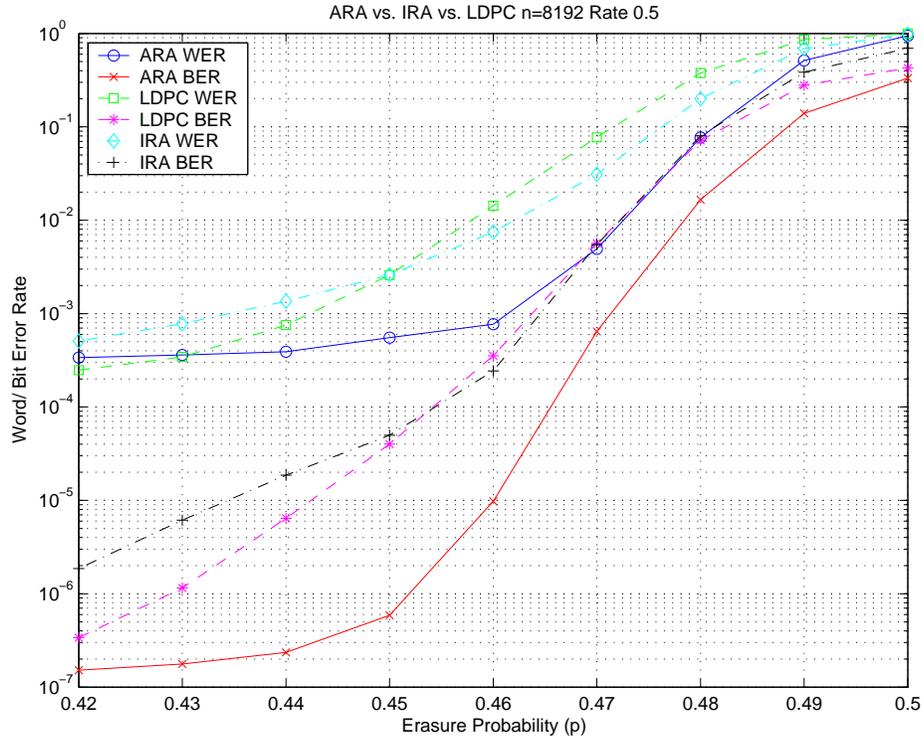

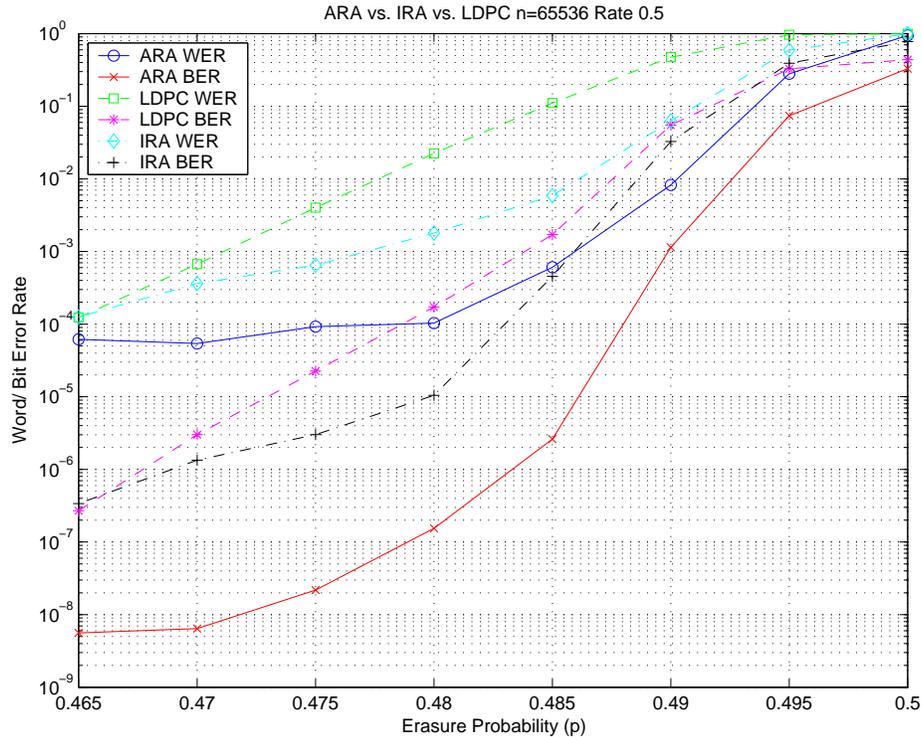

Figure 3: Simulations are shown for the ensembles of ARA and NSIRA codes in Sections 5.1 and 5.2, and right-regular LDPC codes [2]. The plots refer to block lengths of $n = 8192$ and 65536 bits (see upper and lower plots, respectively) and a design rate of 0.5 bits per symbol. Since the ensemble averaged performance is simulated, high-rate outer codes (rates 8179/8192 and 65520/65536, respectively) are used to lower the error floor due to small stopping sets. These outer codes are chosen uniformly at random from the ensemble of the binary linear block codes and their rate loss is neglected.

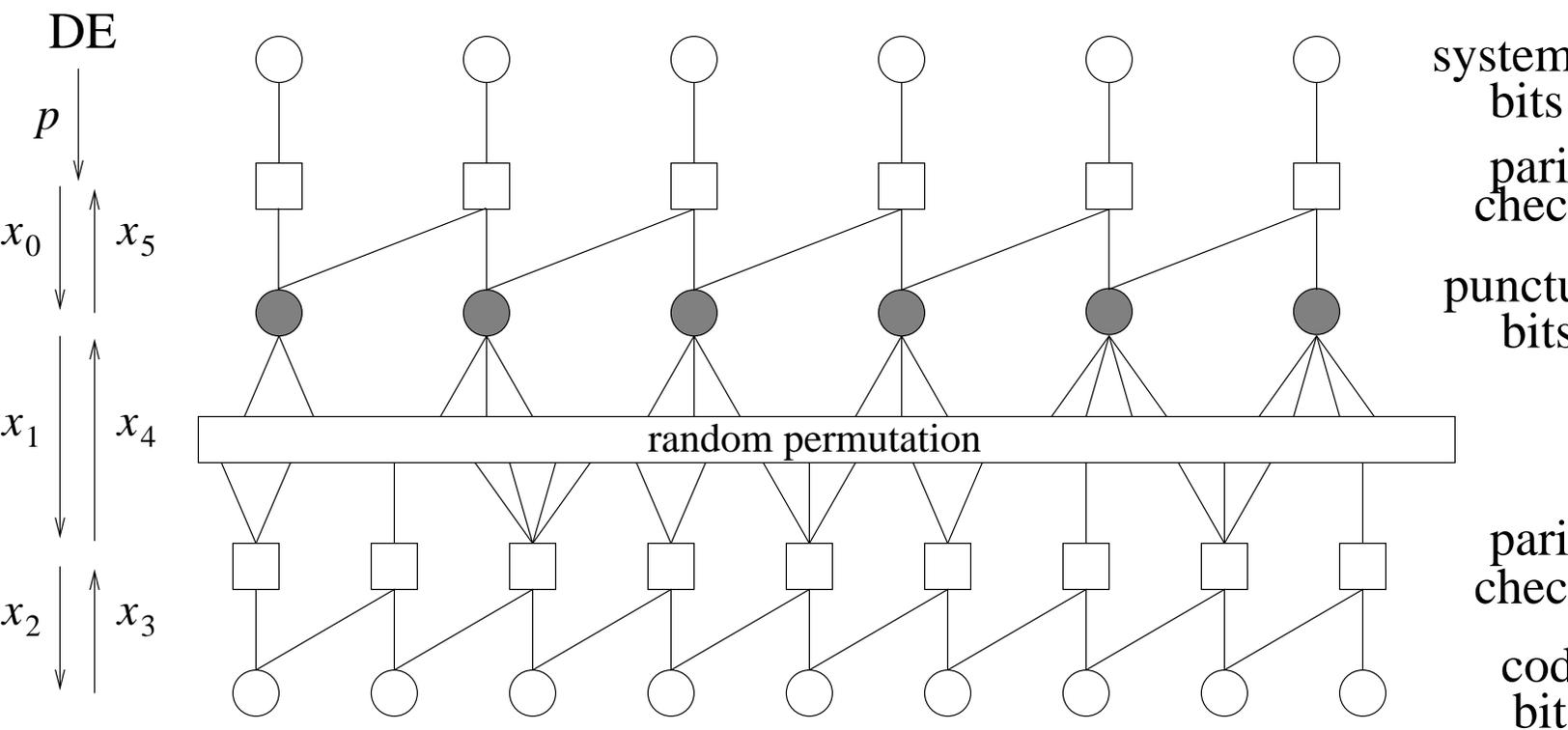